# Realization and manipulation of spiral charge density waves in a two-dimensional metal

Lili Zhou[1,#], Ruizi Zhang[2,#], Chen Si[3,#], Zhaoteng Dong[1,4], Mengya Ren[1], Keru Guo[3], Can Zhang[1,4], Jizheng Wu[3], Fudi Zhou[1], Huixia Yang[4], Yaxin Zhao[4], Guoyuan Yang[1], Xiaolong Xu[4], Yuanxiao Ma[4], Xiao Kong[2,*], Yu Zhang[1,*], Yeliang Wang[1,4,*]

[1]School of Interdisciplinary Science, State Key Laboratory of Environment Characteristics and Effects for Near-space, Beijing Institute of Technology, Beijing 100081, China.

[2]National Key Laboratory of Materials for Integrated Circuits, Shanghai Institute of Microsystem and Information Technology, Chinese Academy of Sciences, Shanghai 200050, China.

[3]School of Materials Science and Engineering, Beihang University, Beijing 100191, China.

[4]School of Integrated Circuits and Electronics, MIIT Key Laboratory for Low-Dimensional Quantum Structure and Devices, Beijing Institute of Technology, Beijing 100081, China.

[#]These authors contributed equally to this work.

[*]Correspondence and requests for materials should be addressed to Yu Zhang (yzhang@bit.edu.cn), Xiao Kong (xkong91@mail.sim.ac.cn), and Yeliang Wang (yeliang.wang@bit.edu.cn).

**Nearly degenerate charge-density-wave (CDW) states play a central role in the competition among collective phenomena. In real materials, however, these states are often intertwined by disorder, hindering their disentanglement and control. Here we show that strain can lift this near-degeneracy and spatially separate distinct CDW states in $NbSe_2$. Using van der Waals (vdW) interactions, we stabilize a micron-scale strain network that produces spatially inhomogeneous strain fields. Within this landscape, the intrinsic 3 × 3 CDW superlattice of pristine $NbSe_2$ transforms into an isolated unidirectional 4 × 1 order under 1D-confined compression, and into a 2 × 2 order under biaxial tension. The 4 × 1 CDW has a multiband origin and exhibits markedly enhanced thermal stability, persisting up to 70 K. At strain-network nodes, it further develops into chiral spiral textures, which can be melted by voltage pulses. These results establish strain as a powerful approach to disentangle, stabilize and manipulate competing electronic orders.**

Understanding how collective electronic order emerges and evolves in quantum materials is a central challenge in condensed-matter physics. Charge density waves (CDWs), characterized by a periodic modulation of electronic density coupled to the lattice[1], provide a paradigmatic example of such order. CDWs are ubiquitous in low-dimensional systems including transition metal dichalcogenides (TMDCs)[2-5], cuprate superconductors[6-8], and kagome lattices[9-11], where they often intertwine with other collective phases such as superconductivity, Mott insulating states, and quantum spin liquids[12-15]. However, unlike the unique CDW instability associated with lattice dimerization in 1D chains, CDW orders in 2D systems can adopt multiple wavevectors and geometries, even within a single compound. These orders are often nearly degenerate in energy, highly susceptible to local perturbations, and capable of generating distinct many-body phenomena. Despite long-standing interest and extensive efforts, disentangling and controlling the CDW orders remain a central challenge.

2H-$NbSe_2$ is a layered TMDC widely regarded as a prototype material for investigating collective electronic orders[16]. Pristine bulk $NbSe_2$ undergoes a nearly 3 × 3 CDW transition upon cooling below $T_{CDW}$ ~ 33 K[16,17], and coexists with superconductivity below $T_C$ ~ 7 K. Notably, this CDW characteristic is highly sensitive to precise atomic arrangements. For example, $T_{CDW}$ exhibits a pronounced layer dependence and is significantly enhanced in the monolayer limit, reaching $T_{CDW}$ ~ 145 K[18-20]. In addition, atomic defects in $NbSe_2$ can serve as pinning centers, locally anchoring the CDW order and even promoting CDW formation above $T_{CDW}$[21-25]. More importantly, multiple CDW orders with triangular, hexagonal, and striped modulations have been observed to coexist in $NbSe_2$, which have been suggested to originate from locally varying strain[26-29]. However, to the best of our knowledge, a consensus on the strain-induced evolution of CDW order in $NbSe_2$ is still lacking.

In this work, we demonstrate that strain can lift the near degeneracy of competing CDW states in $NbSe_2$ and spatially separate distinct CDW orders. Using van der Waals (vdW) interactions, we realize a micron-scale strain network in $NbSe_2$, giving rise to spatially inhomogeneous strain fields. Atomically resolved scanning tunneling microscopy (STM) measurements reveal that the intrinsic 3 × 3 CDW ground state of pristine $NbSe_2$ can be driven into distinct emergent orders: an isolated unidirectional 4 × 1 order under 1D-confined compression, and a 2 × 2 order under biaxial tension. Spectroscopic imaging further uncovers the multiband nature of the 4 × 1 order and identifies that it remains robust even up to 70 K. At the nodes of the strain network, the 4 × 1 order further develops into chiral spiral textures, which can be further melted by voltage pulses. Taken together, these results establish strain as a powerful tuning parameter for disentangling, stabilizing and controlling competing electronic orders.

In our experiments, high-quality $NbSe_2$ single crystals were cleaved under ultrahigh vacuum at room temperature and immediately transferred to the STM head held at 5.0 K (see Methods). The mechanical flexibility of $NbSe_2$, together with its weak interlayer adhesion, naturally favors wrinkle formation in the topmost monolayer during exfoliation[30-32], with the wrinkle density strongly influenced by the tape release rate. Figure 1(a) shows a representative large-scale STM image of the cleaved $NbSe_2$ surface, on which a dense network of interconnected wrinkles is observed. A magnified view in Fig. 1b resolves three characteristic regions: a pristine region (region I), a plateau region atop the wrinkle (region III) whose apparent height is approximately 100 pm greater than that of region I, and an intermediate region (region II) that forms a sloped connection between them.

Figures 1(c)-1(e) show atomically resolved STM images acquired from the regions I, II, and III marked in Fig. 1(b), respectively. In each region, the 1 × 1 atomic lattice is superimposed with a distinct superstructure, specifically triangular 3 × 3 in region I, triangular 2 × 2 in region II, and unidirectional 4 × 1 in region III, as further confirmed by the corresponding Fourier transforms shown in Figs. 1(f)-1(h). The associated wavevectors exhibit almost no energy dispersion (Supplementary Figs. 1-3), supporting their identification as CDW orders. Unlike previous studies suggesting the coexistence of multiple CDW orders in $NbSe_2$[28,29], our results show that the wrinkle can spatially separate these orders, thereby providing an ideal platform to clarify the local conditions that stabilize each order.

Wrinkles in layered 2D materials can be understood as a mechanical response to compressive strain[33], arising from the competition among stretching-energy release, bending-energy cost, and adhesion-energy loss. Our DFT calculations (Supplementary Fig. 4) show that, with increasing compressive strain, the wrinkle morphology evolves from a single-peak structure to a plateau-like profile with an increased height, and eventually into a multi-peak configuration, as schematically illustrated in Fig. 2(a). The formation of a plateau-like profile facilitates the release of compressive strain. However, an excessively large plateau leads to strong adhesion-energy loss, thereby driving the system toward a multi-peak configuration.

Figure 2(b) exhibits the stable atomic structure of an $NbSe_2$ layer under 1% compressive strain, together with the corresponding strain-field distribution. As we can see, the plateau region atop the wrinkle is under compressive strain, the sloping side regions experience tensile strain, and the region away from the wrinkle remains nearly unstrained. Comparing with the measured topography in Fig. 1(b), regions I, II, and III can thus be assigned to the nearly unstrained, tensile-strained and compressive-strained regimes, respectively, which is also consistent with previous Raman scattering results[34].

To clarify the relationship between the experimentally observed CDW orders and local strain, we first calculate the strain-dependent phonon instability of $NbSe_2$, which is widely regarded as a direct fingerprint of CDW formation. In pristine $NbSe_2$, the prominent soft-phonon mode near 2/3 ΓM indicates the emergence of the $3 \times 3$ CDW. Biaxial tensile strain shifts the soft-phonon mode from 2/3 ΓM towards ΓM, suggesting an evolution of the ground state from the $3 \times 3$ to $2 \times 2$ CDW. By contrast, compressive strain drives the soft-phonon mode towards 1/2 ΓM, favoring a unidirectional $4 \times 1$ CDW (Fig. 2(c)). To further evaluate the relative CDW stability, we calculate the strain-dependent energy landscape of the three candidate CDW orders. As shown in Fig. 2(d), the $4 \times 1$ CDW is stabilized under uniaxial compressive strain, whereas the $2 \times 2$ CDW becomes energetically preferred under biaxial tensile strain exceeding 2.5%. In the intermediate tensile-strain regime (0~2.5%), the $3 \times 3$ and $4 \times 1$ CDWs are nearly degenerate in energy, indicating the possibility of their coexistence. This strain-selective CDW evolution is consistent with the experimental observations in Figs. 1(c)-(e) and can be further corroborated by the simulated STM images shown in

Supplementary Fig. 5.

In previous studies, the $4 \times 1$ CDW in $NbSe_2$ was found to coexist and intertwine with the $3 \times 3$ CDW without an obvious distinction in STM topography. Here, by exploiting $NbSe_2$ wrinkles, we engineer a 1D-confined compressive strain that is laterally clamped by tensile strain. This strain geometry lifts the near-degeneracy between the competing CDW states and spatially disentangles them, thereby providing an unprecedented opportunity to controllable stabilization of an isolated $4 \times 1$ CDW.

A central aim of this work is to elucidate the formation, characteristics, and evolution of this newly identified isolated unidirectional $4 \times 1$ CDW. By examining dozens of cases, we find that this $4 \times 1$ CDW emerges only when the wrinkle height exceeds ~50 pm (Supplementary Fig. 6). Notably, the $4 \times 1$ CDW stripes consistently align along the high-symmetry lattice directions while remaining tilted with respect to the wrinkle edges. As the wrinkle edges are approximately aligned with lattice directions[35], this indicates that the CDW stripes preferentially develop along one of the remaining crystallographic orientations.

To rationalize this behaviour, we calculate the energy of the $4 \times 1$ CDW for different stripe configurations under varying compressive strains. As shown in Fig. 2(e), stripes oriented at 60° relative to the strain direction are consistently lower in energy than those aligned parallel, indicating a robust energetic preference for the rotated orientation. This preference arises because the 60°-rotated configuration minimizes internal strain during the phase transition under applied compressive strain. As summarized in Supplementary Table 1, the lattice constant of $4 \times 1$ CDW is slightly larger and smaller than $3 \times 3$ CDW along the armchair and zigzag direction, respectively. In this geometry, the lattice constant of the 60°-rotated configuration changes less under the same applied strain, thereby reducing the internal strain and overall energy cost. Furthermore, when the wrinkle width exceeds ~6 nm, the two stripe orientations corresponding to the two remaining crystallographic orientations can coexist, as exhibited in Supplementary Fig. 7.

We next examine the stability of the isolated unidirectional $4 \times 1$ CDW. Temperature-dependent STM measurements shown in Figs. 2(f)-(i) reveal that the $4 \times 1$

CDW remains robust, with no appreciable attenuation even up to 70 K (the upper limit of our experimental conditions). This behaviour contrasts sharply with the intrinsic $3 \times 3$ CDW in pristine $NbSe_2$, where the CDW is almost suppressed at ~33 K (Supplementary Fig. 8). Theoretically, the CDW transition temperature is positively correlated with $\Delta E$, where $\Delta E$ denotes the energy difference between the CDW state and its pristine counterpart[36,37]. Our calculations show that $\Delta E$ is about 3.5 meV per formula unit in the absence of strain for the $3 \times 3$ CDW, whereas it reaches to ~7.4 meV per formula unit under 3% compressive strain for the $4 \times 1$ CDW. These results suggest that the $4 \times 1$ CDW has a higher transition temperature.

Strikingly, at the nodes of the strain network, unidirectional $4 \times 1$ CDWs propagate along wrinkles and converge, with stripes orientations differing by 60° or equivalently 120°. Geometric constraints then drive the three $4 \times 1$ CDWs to evolve into a chirality spiral configuration, as exemplified in Fig. 3(a). Figures 3(b)-(d) present corresponding bias-dependent spectroscopic maps. In contrast to conventional CDW, where out-of-phase charge modulations are typically observed on opposite sides of the CDW gap, our measurements reveal a more complex spatial evolution in phase. To quantify this behaviour, we extract line profiles of the dI/dV intensity as a function of sample bias shown in Fig. 3(f). These results remind us of the presence of an additional charge modulation, associated with a gap opening in a distinct electronic band away from the Fermi level $E_F$, thereby giving rise to the observed bias-dependent phase evolution, as schematically illustrated in Fig. 3(g).

Our measured STS spectra shown in Fig. 3(e) further confirm the presence of two CDW gap signatures at the sample biases of approximately -0.44 V and 0.01 V, respectively. The low-energy bands of $NbSe_2$ are dominated by the bonding and antibonding combinations of Nb-4*d* orbitals, arising from interlayer coupling. As demonstrated in previous studies[38], the two harmonic charge modulations associated with these bands exhibit a relative phase shift that can deviate from π. The precise bias dependence is governed by the magnitudes of the two gaps, the gap energies relative to the $E_F$, and the relative phase shift between the two charge modulations[39]. Therefore, our observations provide direct evidence for the two-band nature of the $4 \times 1$ CDW.

Despite its thermal robustness, the unidirectional 4 × 1 CDW can be controllably driven into a melted state by tip voltage pulses. Figures 4(a) and (b) show representative STM images acquired before and after a pulse, respectively. After applying a pulse of V = 5 V with a duration of 10 ms, the unidirectional 4 × 1 CDW (Fig. 4(c)) loses its long-range order, retaining only short-range correlations. The initially continuous stripe pattern with long-range periodicity consequently breaks into spatially disconnected segments with a markedly reduced correlation length, ultimately evolving into fragmented, labyrinthine patches (Fig. 4(d)). Notably, throughout this process, the wrinkle morphology is essentially unchanged and the atomic lattice remains intact. By contrast, the 3 × 3 CDW in region I and the 2 × 2 CDW in region II are hardly affected under the same pulse conditions (Supplementary Fig. 9).

Such voltage-pulse-induced CDW melting is further substantiated by Fourier transform analysis. Prior to the pulse, the unidirectional 4 × 1 CDW superlattice gives rise to two sharp peaks at the $4a_0$ wavevectors in the Fourier image, as enclosed by the red circles in Figs. 4(e). Upon application of the pulse, these superlattice peaks become azimuthally broadened, evolving into a weak and diffuse ring that progressively merges into the background (Figs. 4(f)), while the Bragg peaks remain well defined. This selective loss of long-range CDW coherence, concomitant with the preservation of the underlying lattice order, is a hallmark of CDW melting[40], as schematically illustrated in Figs. 4(g) and (h).

CDW melting has traditionally been achieved by increasing temperature, which enhances thermal fluctuations and progressively suppresses both the amplitude and phase coherence of the CDW order parameter. Under equilibrium conditions, this process is typically accompanied by structural transformations, proceeding from an ordered CDW state through hexatic and nematic intermediates into a fully disordered state[41,42]. More recently, nonequilibrium pathways to CDW melting have been reported via ultrafast optical excitation, where the system is driven into a transient metastable state with picosecond lifetimes, providing a route to disentangle electronic instabilities from electron-phonon coupling[43-45]. In contrast, field-driven CDW melting remains largely unexplored. Here, we demonstrate that a voltage pulse alone can also induce

CDW melting. This process is essentially instantaneous and can be viewed as an equilibrium pathway from a long-range ordered CDW into a state comprising spatially disconnected CDW segments, reminiscent of the theoretically proposed vestigial order[46]. The ability to locally induce CDW melting points to a mechanism distinct from thermally and photoinduced transitions, which are typically governed by the proliferation of topological defects and the gradual loss of long-range order[47,48].

In summary, we construct a micron-scale strain network in $NbSe_2$ that lifts the near degeneracy of competing CDW states and spatially separates distinct CDW orders, thus enabling direct visualization of their atomic-scale evolution. We show that the intrinsic $3 \times 3$ CDW order of pristine $NbSe_2$ can be driven into an isolated unidirectional $4 \times 1$ order under 1D-confined compression, and into a $2 \times 2$ order under biaxial tension. The $4 \times 1$ CDW has a multiband origin and exhibits markedly enhanced thermal stability. At strain-network nodes, it further develops into chiral spiral CDW textures, which can be melted by voltage pulses. These results establish strain as a powerful approach to disentangle, stabilize, and manipulate competing electronic orders.

## Methods

### STM/STS measurements.

STM/STS measurements were performed using a commercial low-temperature system (Createc) operated at 4.7 K with a base pressure better than $1 \times 10^{-10}$ mbar. The STM images were taken in a constant-current scanning mode. An electrochemically etched tungsten tip was used as the STM probe, which was calibrated by using a standard graphene lattice and a Si(111)-(7×7) lattice. The STS measurements were taken with a standard lock-in technique by turning off the feedback circuit and using a 793-Hz 5 mV A.C. modulation of the sample voltage.

### DFT calculations.

The calculations are carried out by using Vienna ab initio simulation package (VASP) with the projector augmented wave (PAW) method[49,50]. A generalized gradient approximation (GGA) in the form of Perdew-Burke-Ernzerhof (PBE) is adopted for the exchange-correlation functional[51]. The wave functions are expanded using a planewave basis set with an energy cutoff of 600 eV and $9 \times 9 \times 1$ and $16 \times 2 \times 1$ K-point sampling was adopted for $3 \times 3$ and $4 \times 1$ CDW phases. The vacuum layer is 20 Å. The atomic configurations are fully relaxed until the residual forces on each atom are smaller than 0.01 eV/Å. Phonon calculations were performed using the finite displacement method as implemented in phonopy code[52]. All atomic structures are displayed with VESTA[53].

### Data Availability

All raw data generated during the current study are available from the corresponding authors upon request.

### Acknowledgments

Y.Z. acknowledges the National Key R&D Program of China (Grant Nos. 2022YFA1402502 and 2022YFA1402602), National Natural Science Foundation of China (Grant Nos. 92580103 and 12274026), Young Elite Scientists Sponsorship

Program of the Beijing High Innovation Plan (Grant No. 20250836), Open Fund of Key Laboratory of Multiscale Spin Physics (Ministry of Education), Beijing Normal University (Grant No. SPIN2025K01), and Beijing Institute of Technology Kunpeng&Ascend Center of Cultivation. Y.L.W. acknowledges the National Key R&D Program of China (Grant No. 2021YFA1400103) and National Natural Science Foundation of China (Grant No. 12321004). X. K. acknowledges the National Natural Science Foundation of China (Grant No. 62304235).

## Author Contributions

Y.Z. and Y.W. coordinated the research project. L.Z., Z.D., M.R., C.Z., F.Z., H.Y., Y.Z., X.X., and Y.M synthesized the samples and performed the STM experiments. R.Z., C.S., K.G., J.W., G.Y., and X.K. performed the theoretical calculations. Y.Z. and X.K. contributed to the overall scientific interpretation and edited the manuscript. All authors were involved in discussions of this work.

## Competing interests

The authors declare no competing interest.

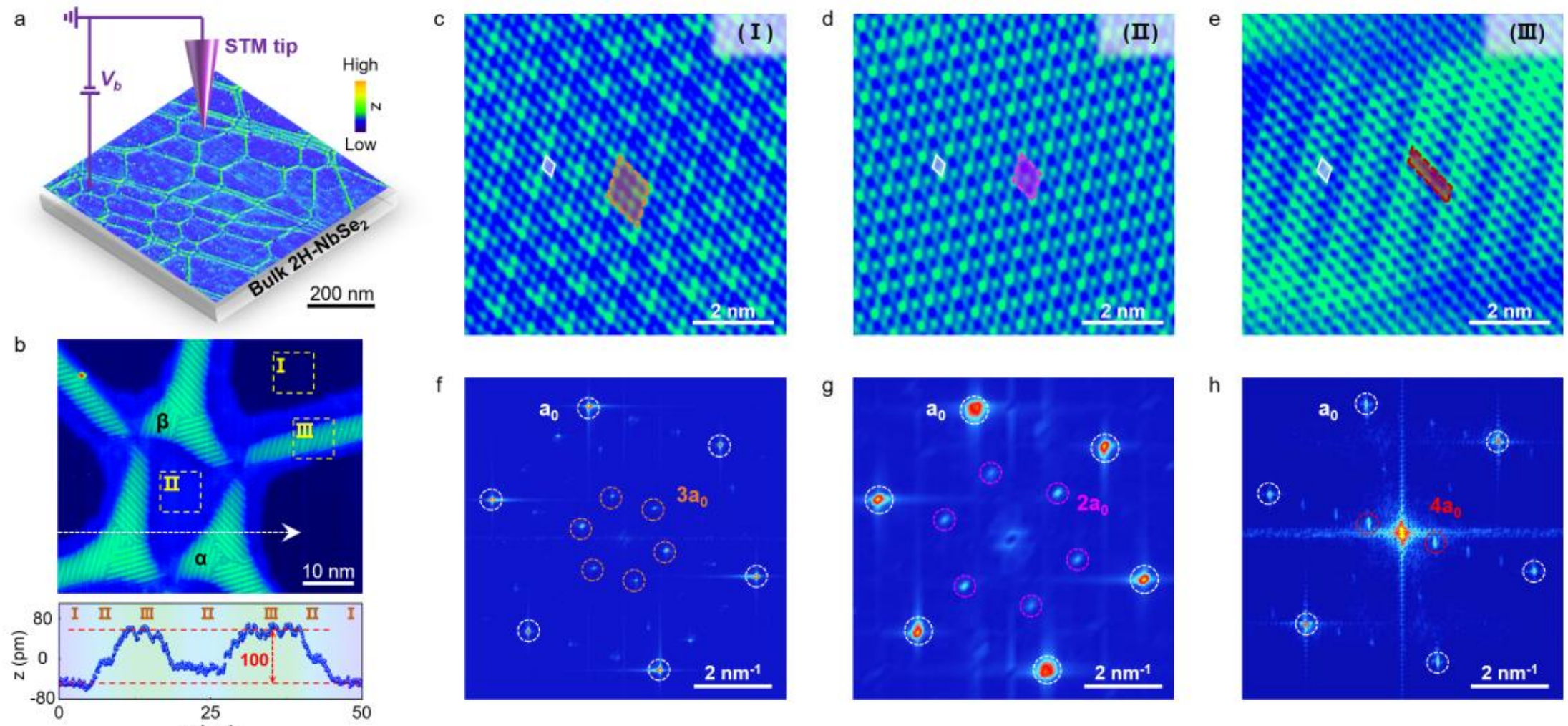

**Fig. 1 | CDW evolution in wrinkled $NbSe_2$. a,** Experimental setup and large-scale topography STM image of cleaved $NbSe_2$ surface, showing a high density of wrinkle networks. Scanning settings: $V_b$ = -1.0 V, $I_t$ = 10 pA. **b,** Representative STM image of $NbSe_2$ acquired at a wrinkle junction, with the line profile recorded along the white arrow. Scanning settings: $V_b$ = -1.0 V, $I_t$ = 10 pA. **c-e,** Atomically resolved STM images recorded at the regions I, II, and III marked in panel b. Scanning settings: $V_b$ = -0.3 V, $I_t$ = 100 pA. **f-h,** Fourier transforms of panels c-e. The outer six bright spots highlighted by the white circles indicate the atomic Bragg peaks ($a_0 \times a_0$) of $NbSe_2$. The bright spots enclosed by the orange, pink, and red circles are related to the CDW orders with the $3a_0 \times 3a_0$, $2a_0 \times 2a_0$, and unidirectional $4a_0$ wavevectors, respectively.

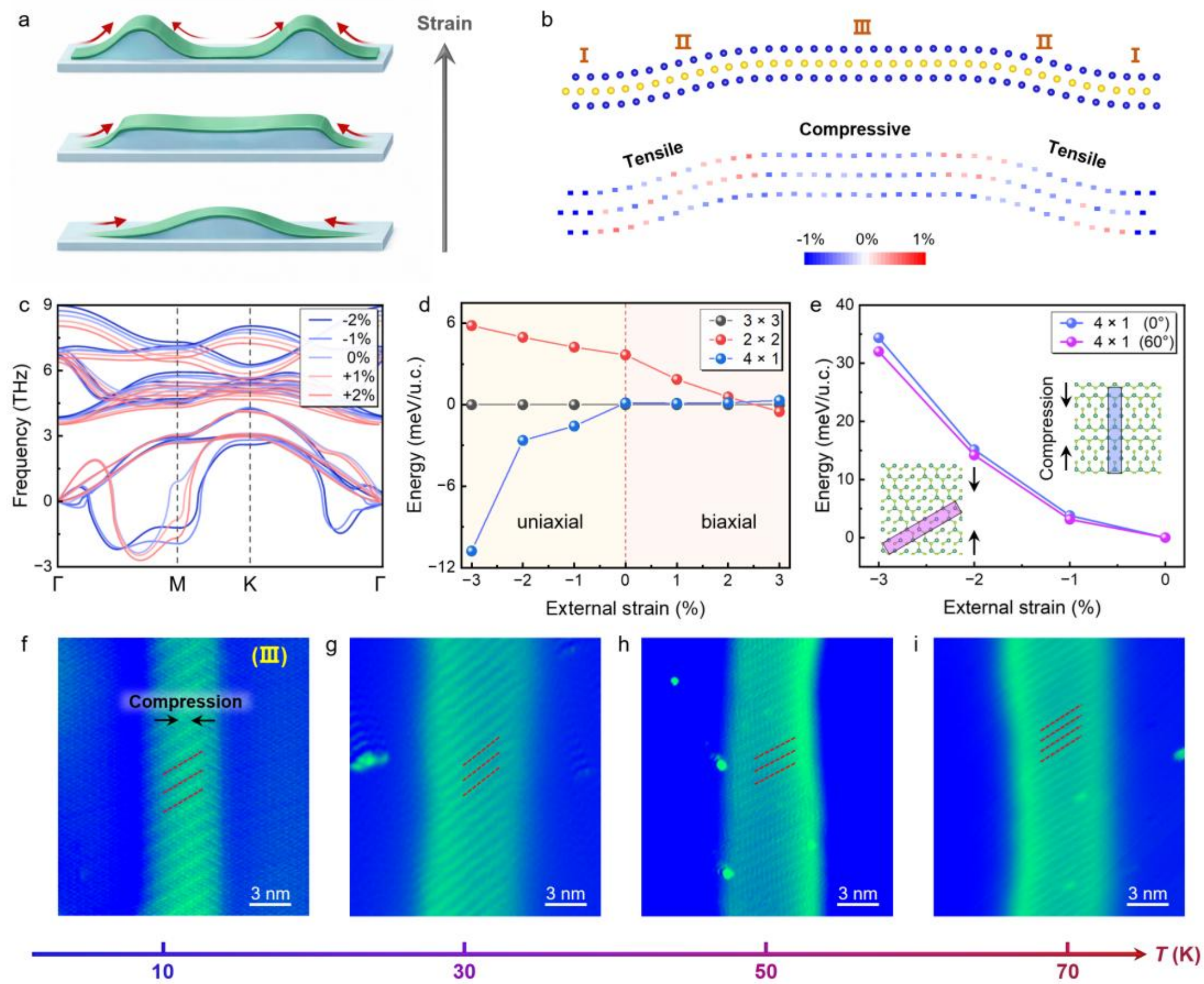

**Fig. 2 | Origin and characteristics of the unidirectional 4 × 1 CDW in $NbSe_2$. (a)** Schematic illustration of the evolution of wrinkle configurations under compressive strain. As the compressive strain increases, the wrinkle morphology gradually evolves from a single-peak wrinkle to a plateau-like wrinkle, and eventually to a multi-peak wrinkle. **(b)** Theoretically calculated relaxed atomic model and strain-field distribution of an $NbSe_2$ wrinkle under 1% compressive strain. The central plateau region is under compressive strain, whereas the sloped side regions experience tensile strain. The region away from the wrinkle remains strain-free. **(c)** Calculated phonon spectra of $NbSe_2$ under different strains. **(d)** Energy differences of the 2 × 2 and 4 × 1 CDW relative to the 3 × 3 CDW as a function of strain. The energy of the 3 × 3 CDW is taken as zero. Here the tensile strain is biaxial, whereas the compressive strain is uniaxial along the armchair direction. **(e)** Energy of the 4 × 1 CDW for different configurations under different strain conditions. Blue and pink spheres represent unidirectional stripes parallel to the strain direction and oriented 60° to it, respectively. **(f-i)** Temperature-dependent STM images of the unidirectional 4 × 1 CDW. The order persists up to 70 K.

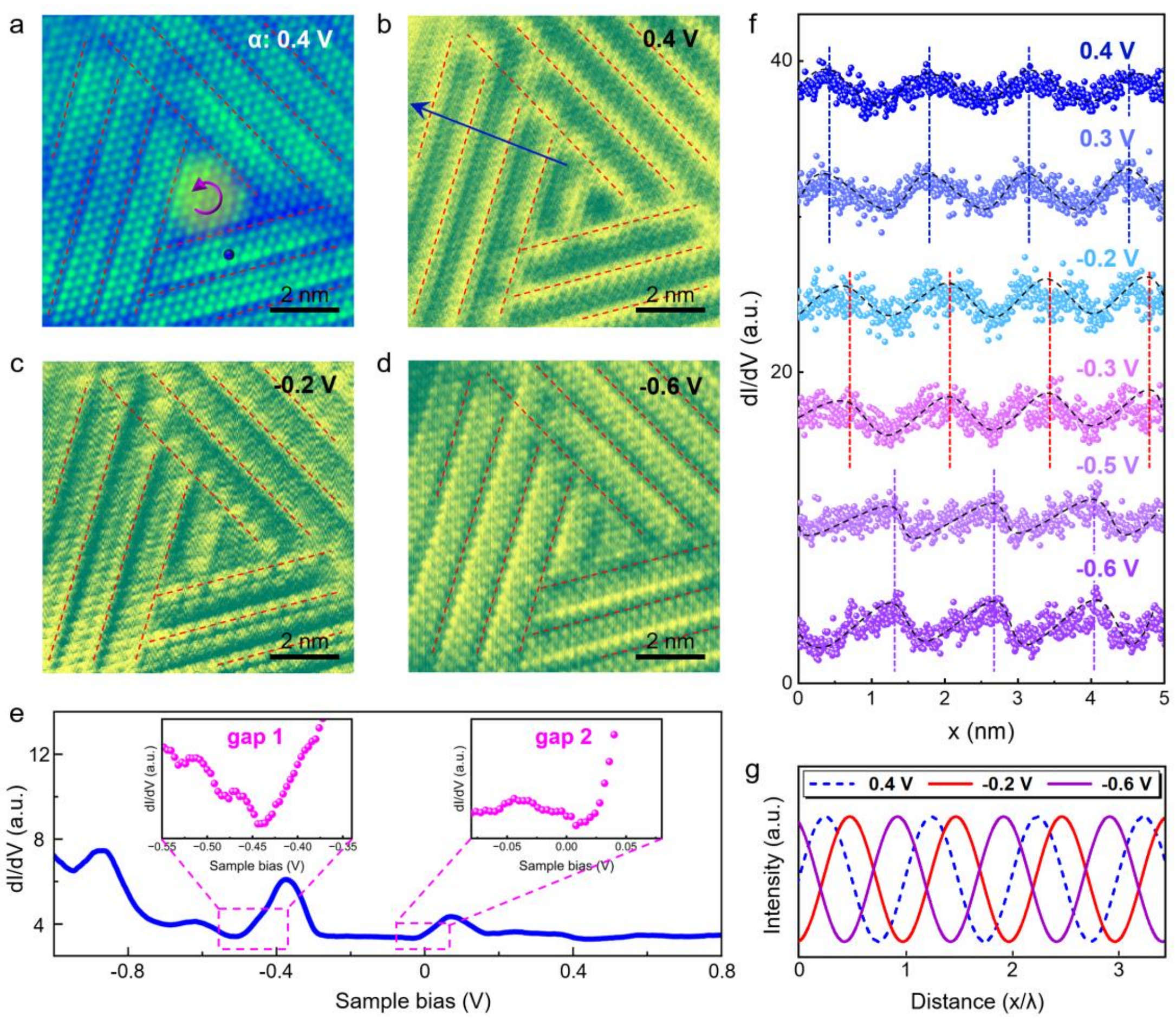

**Fig. 3 | Multiband and chiral nature of the unidirectional CDW in $NbSe_2$. (a)** Representative STM image acquired in region α marked in Fig. 1(b). The unidirectional CDW propagating along three crystallographic directions converge at a point, giving rise to a well-defined chirality, as indicated by the central arrow. The dashed lines indicate the periodicity of the unidirectional CDW modulation, as determined by visual guidance. **(b-d)** Spectroscopic maps recorded at the same region as panel a under different sample biases. The dashed lines indicate identical spatial locations. **(e)** Typical STS spectrum recorded at the blue dot in panel a. The inset shows zoom-in STS spectra, highlighting the two gap features. **(f)** Bias-dependent dI/dV height profiles acquired along the blue arrow marked in panel d. The curves are vertically offset for clarity. The vertical dashed lines mark the positions of the height maxima. **(g)** Schematic bias-dependent charge density modulations extracted from panel g. The dashed and solid lines correspond to negative and positive sample biases, respectively, and exhibit a non-π/2 phase difference, which can be well captured by multiband CDW physics.

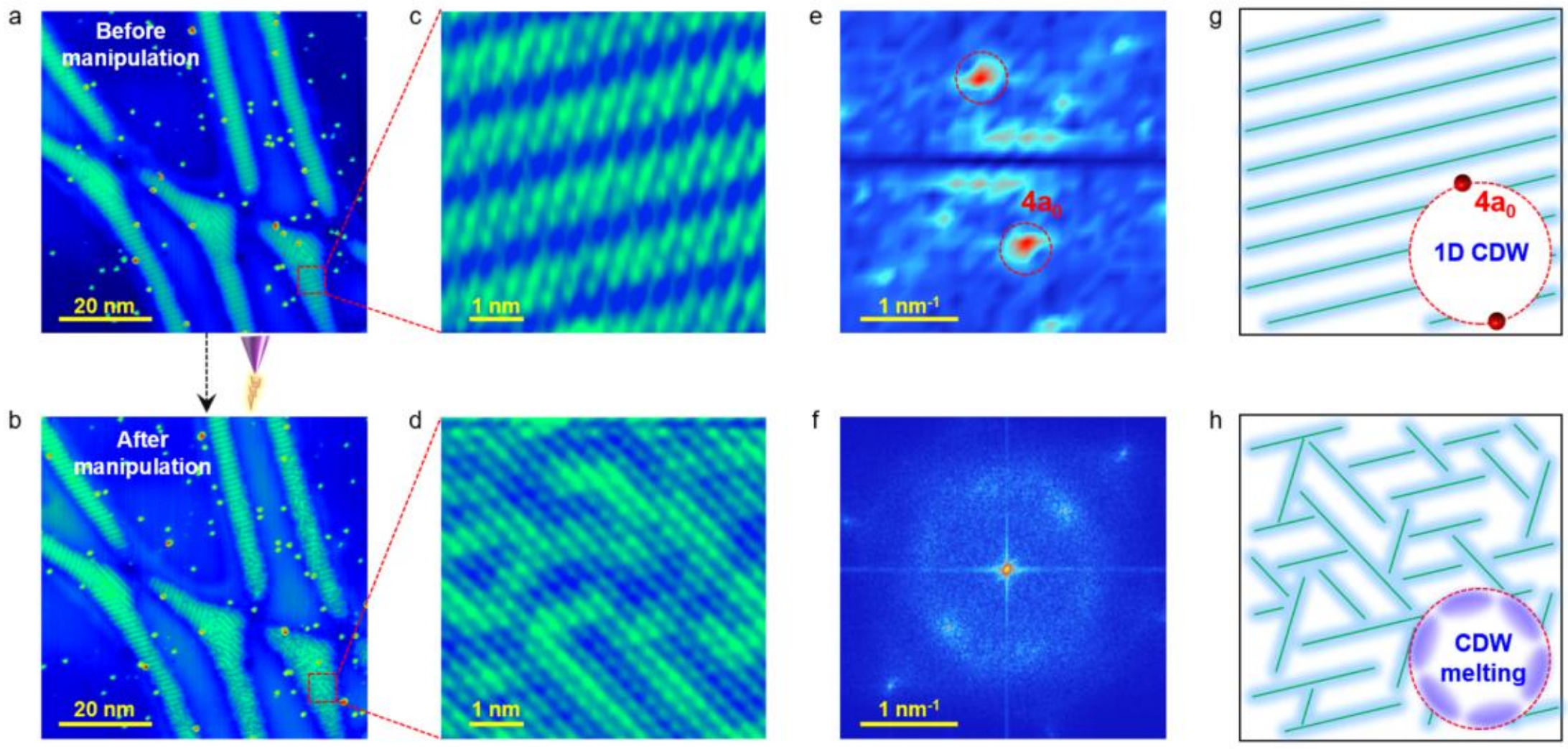

**Fig. 4 | Melting of the unidirectional CDW in $NbSe_2$. (a,b)** Representative STM images of $NbSe_2$ acquired before and after applying a STM tip pulse, respectively. **(c,d)** Atomically resolved STM images taken from the regions marked by the dashed red rectangles in panels a and b, respectively. **(e)** Fourier transform of panel c. The bright spots highlighted by the red circles indicate the unidirectional $4a_0$ CDW order. **(f)** Fourier transform of panel d, showing reduced intensity and peak broadening at the $4a_0$ wavevector. **(g,h)** Schematic illustrations of the unidirectional $4a_0$ CDW and its melting process.